\documentstyle[11pt,newpasp,twoside,epsf]{article}
\begin{document}
\title{HST images of B2 radio galaxies: the link between circum-nuclear dust and radio properties}
\author{H. R. de Ruiter}
\affil{Osservatorio Astronomico di Bologna, Via Ranzani 1, I-40127 Bologna, Italy}
\author{P. Parma, R. Fanti}
\affil{Istituto di Radioastronomia del CNR, Via Gobetti 101, I-40129 Bologna, Italy}
\author{A. Capetti}
\affil{Osservatorio Astronomico di Torino, Strada Osservatorio 25, I-10025 Pino Torinese, Italy}
\author{R. Morganti}
\affil{Netherlands Foundation for Research in Astronomy, Postbus 2, NL-7990 AA Dwingeloo, the Netherlands}
\begin{abstract}

\end{abstract}
\section{Introduction}
A long standing question in radio astronomy has been why some elliptical galaxies
host strong radio sources and others not. The reason may have something to do with the
physical conditions near the (active) nucleus from which the radio source emanates.
The resolution of HST images is such that dust features close to the nucleus can be 
detected, and these may often be in the form of circum-nuclear disks. A study of dust
features in 3C radio galaxies and their relation to the radio source properties was 
presented by de Koff et al. (2000). The large majority of the 3C radio
galaxies are of the Fanaroff-Riley type II, i.e. they have high radio 
luminosities.

A number of lower luminosity (FR I) radio sources from the B2 sample (see e.g. Fanti et al. 1987) have been imaged recently in two colors ($V$ and $I$) with the WFPC2 of the Hubble Space Telescope
(Capetti et al. 2000). Although only about 60 \% of the B2 sample has been
observed, this should not have introduced any selection bias (the selection for observation
was in practice done in random fashion).

In order to study the influence of dust on the radio sources we produced images which show
only regions with dust (if present). These absorption maps were obtained by first constructing 
a model of the
brightness profile of the galaxy (in $I$), which was then, properly scaled, divided by
the $V$ image of the galaxy. The galaxy was thus removed, while small scale absorption
features caused by the presence of dust are then easily visible. 

Of the B2 sources observed with the HST slightly more than half (53~\%) show signs of dust,
in most cases in the form of lanes or disks. Only in a minority the dust features are more 
complicated (e.g. multiple distorted filaments or warped lanes). Interestingly, if the dust
structure is complicated the radio source tends to have no detected radio jet (seven out of nine),
while simple dust structures (lanes or disks) are often associated with radio sources with
jets (thirteen out of twenty). There is no significant difference in radio power and redshift
between sources with and without dust, it is therefore unlikely that there are strong selection
effects that might affect our analysis.

We use a Hubble constant of 100 km~s$^{-1}$Mpc$^{-1}$. 

\section{Radio-dust alignments}
We have many examples in which the dust feature and the radio jet direction appear to be close to
perpendicular. This type of alignment has been known since the seventies (Kotanyi \& Ekers 1979). According to de Koff et al. (2000), who used very recent HST images of 3C sources,  
this alignment is most pronounced in FRI-type sources, while the disk structures are usually very small.
In FRII sources they again find alignment if the dust is in small disks, but otherwise there is no strong
evidence for alignment.

Here we enlarge on the discussion given by de Koff et al. (2000), and amend some of their conclusions
concerning FRI sources. 
\begin{figure}
\plottwo{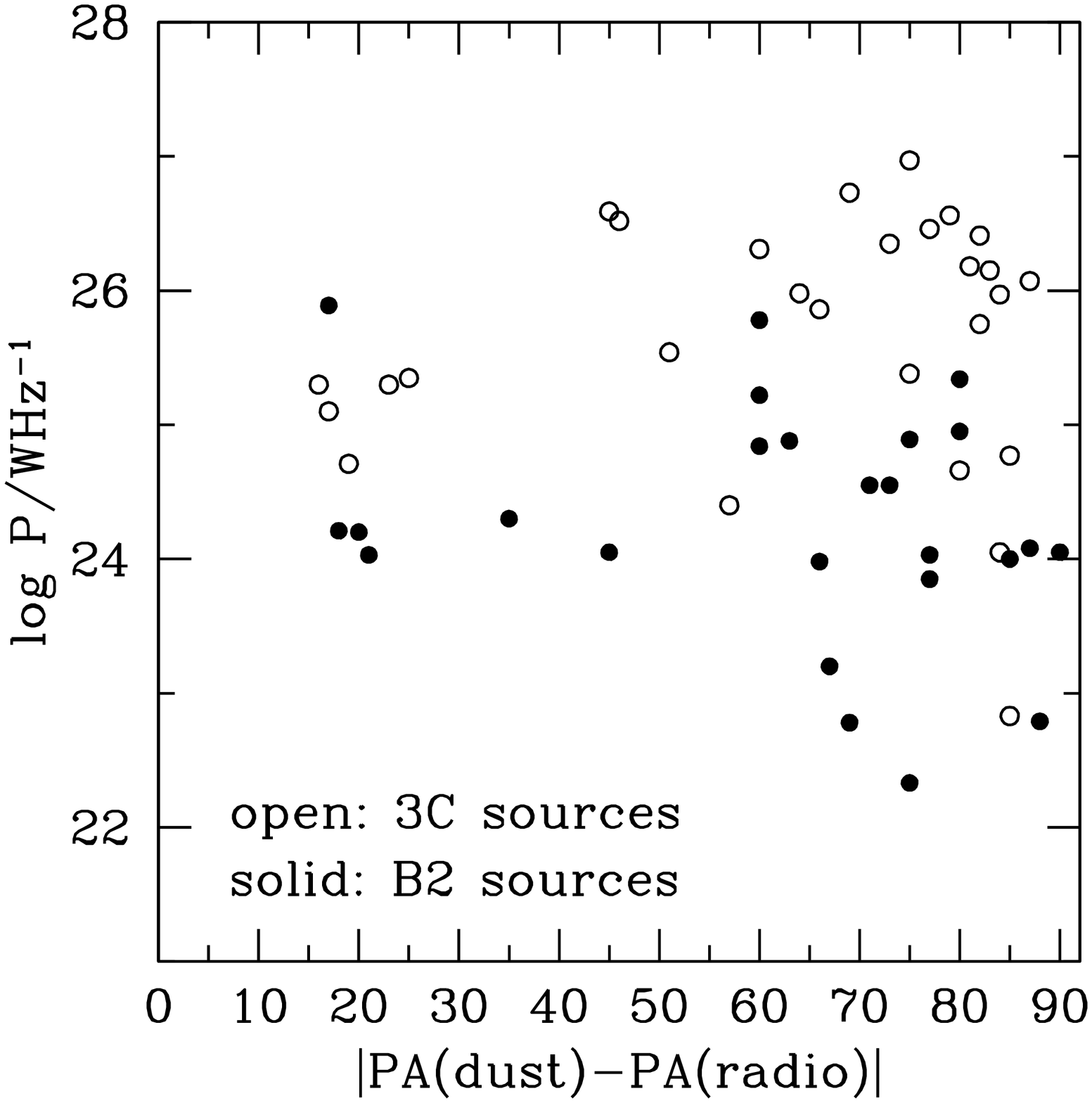}{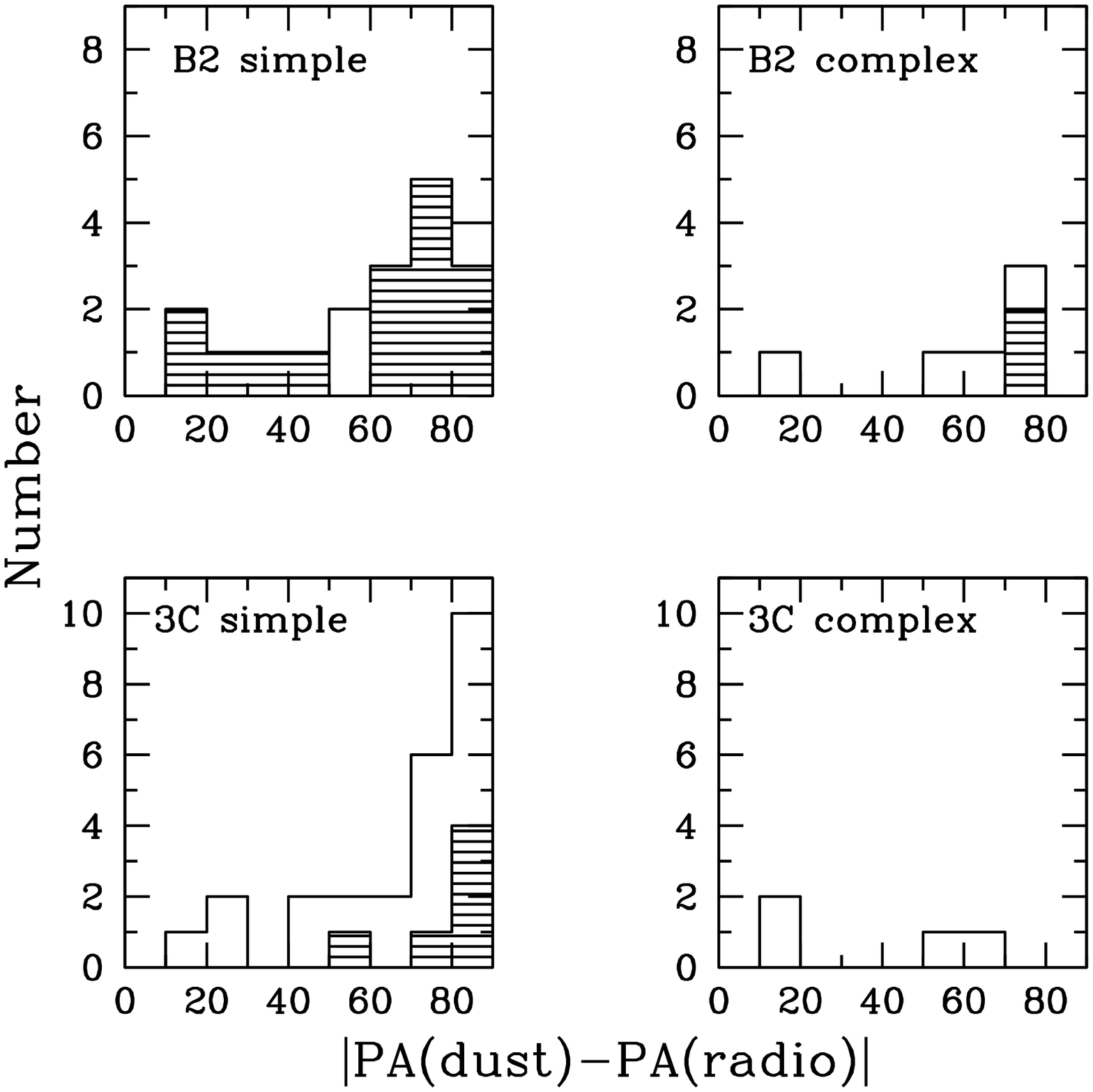}
\caption{The position angle difference between dust feature and radio jet as a function of radio power (left) and its distribution in B2 and 3C sources, according to the complexity of the dust structure (right)}
\end{figure}
First of all, if we plot the difference
in position angle between the dust feature and the radio source (see Fig.~1, left), we notice that there is a large spread in
position angle for FRII sources, while FRIs are indeed crowded towards $\Delta$(P.A.)$\sim 90^{\rm o}$, {\it provided
that the luminosity is below} $10^{24}$~WHz$^{-1}$. The stronger tail of FRI sources appears to behave
more like FRIIs.
Although the statistics are poor the alignment appears to
be more convincing for sources with a simple dust structure (that is single lines or obvious disklike
features). This is shown in Fig.~1 (right). 

The shaded part of the histogram gives the contribution
of FRI sources.
 
\section{Dust masses and radio power}
We calculated the dust masses in the usual way (see Sadler \& Gerhard 1985), making
use of the package SYNAGE++, recently developed at the Istituto di Radioastronomia (Bologna) by
Dr. M. Murgia. We determined the covering factor of the dust by summing all pixels in the absorption maps
with a value $<0.85$, following de Koff et al. (2000).
Most striking is the approximately linear correlation between radio power and dust mass, for the sources in
which dust was detected. This is shown in Fig.~2  (left), in which we have plotted B2 and 3C sources in which dust features are seen. 
\begin{figure}
\plottwo{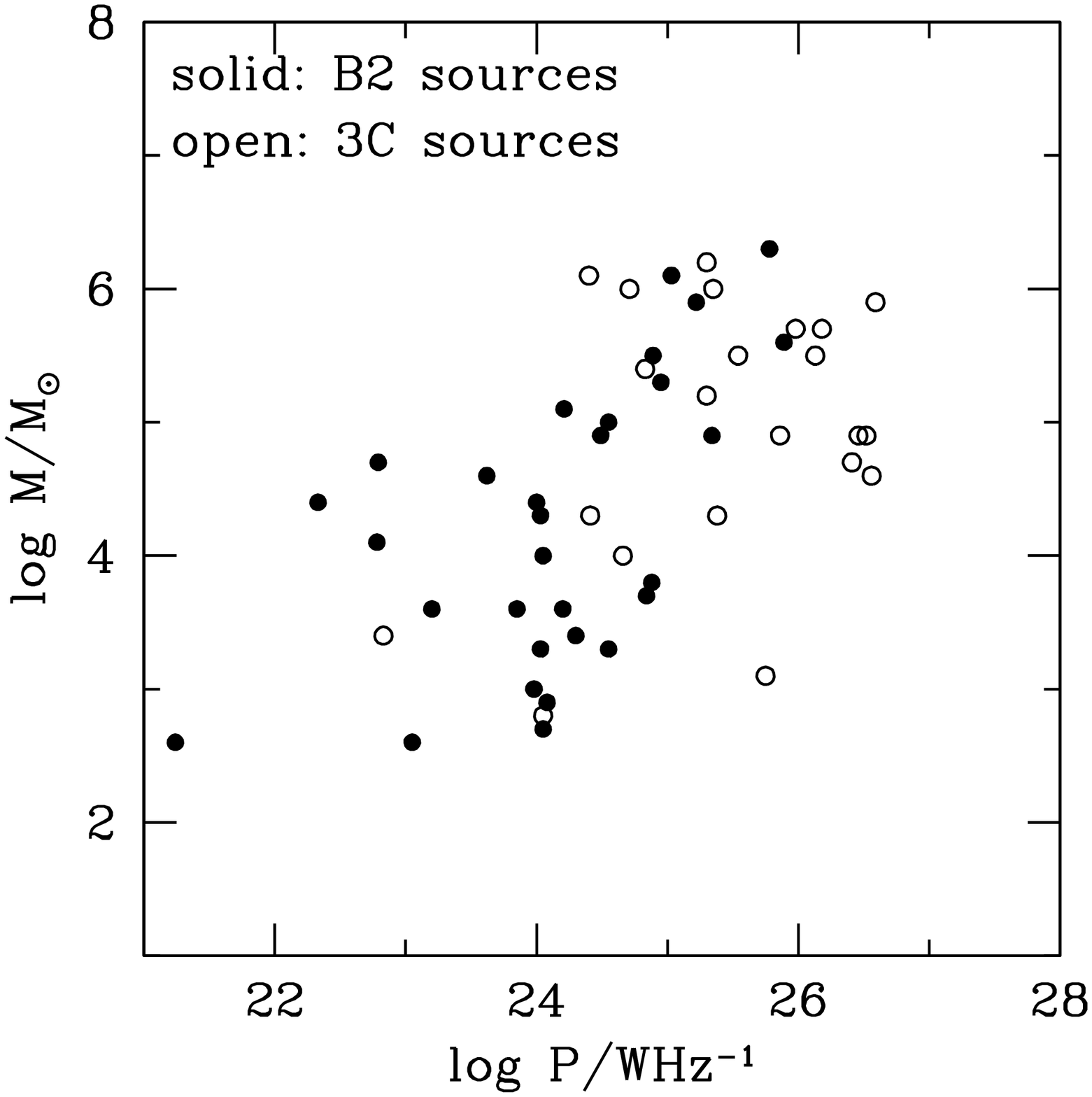}{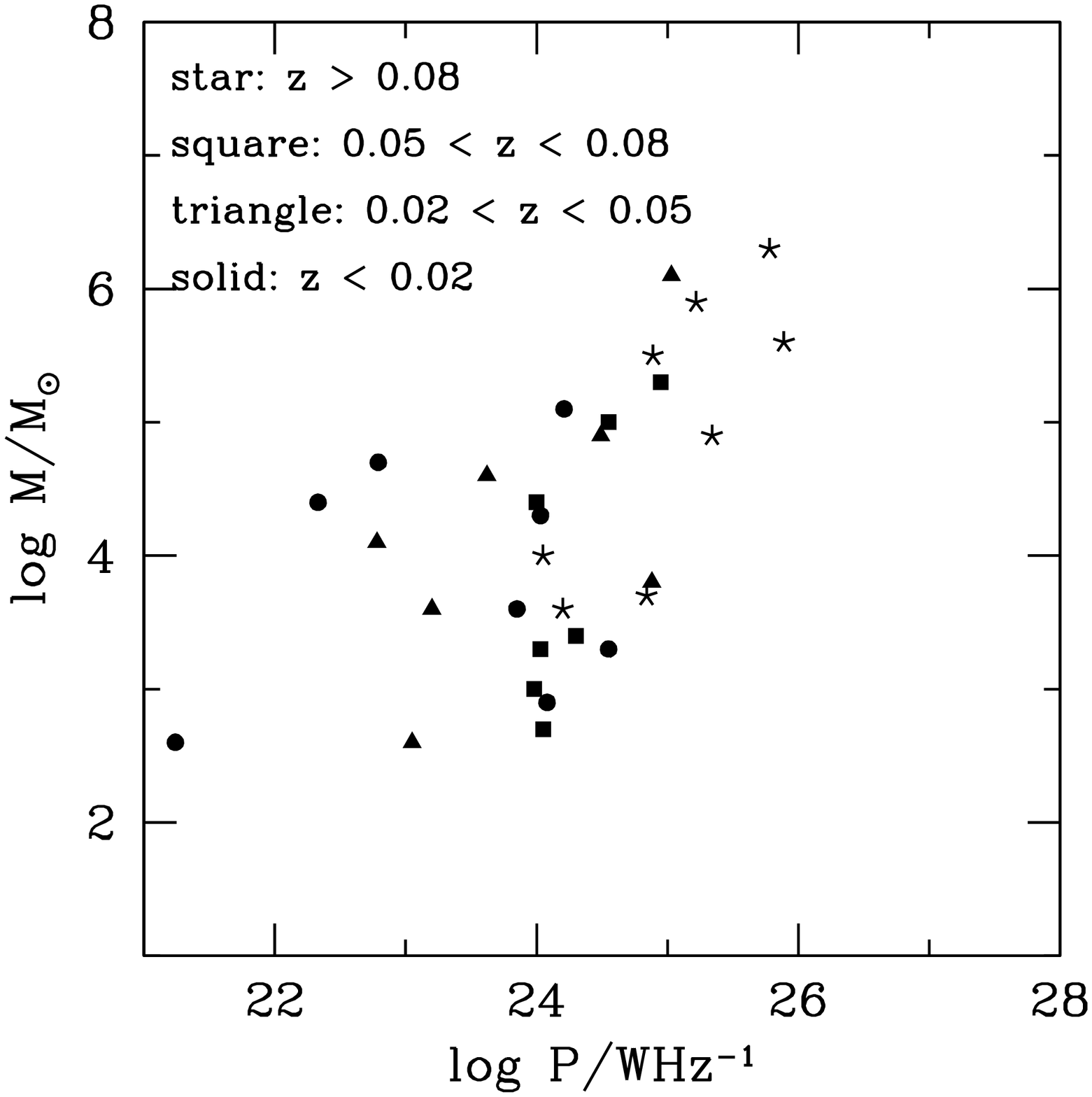}
\caption{The relation between (total) radio power and dust masses. Left: for B2 and 3C sources; right: B2 sources in different redshift intervals}
\end{figure}

Although there are many upper limits to dust masses (not shown in the figure)
these most likely do not destroy the correlation.  A similar correlation holds if instead of the total radio power
we consider the radio power of the core. In Fig.~2 (right) we have
divided the B2 sources into four redshift intervals. It is quite evident that the correlation remains valid inside
all redshift intervals, and we can therefore conclude that it is not due to a selection
effect depending on redshift.

There is a hint that the alignment is better if the dust mass is smaller; 
such dust features typically are $< 500$~pc.
This may be taken as supporting evidence for 
the statement of de Koff et al. (2000), that smaller dust features tend to be better aligned.
For sources with $\Delta (P.A.) < 60^{\deg}$ we find $M/M_{\sun} = 10^{5.6{\pm 0.3}}$,
while well aligned sources with $\Delta (P.A.) > 60^{\deg}$ have $M/M_{\sun} = 10^{4.4{\pm 0.3}}$. This
difference is due exclusively to 3C sources, because B2 alone (i.e. mostly FRI sources), have a median of 
$M/M_{\sun} = 10^{4.4}$, irrespective of alignment angle. It is clear anyway that the lower masses at angles 
$> 60^{\deg}$ are mostly associated with FRI sources with power $P$ below $10^{24}$WHz$^{-1}$, in agreement with
what was said in Section 2. 


\section{Conclusions}
We have shown that dust (specifically its morphology and amount) 
is linked with the properties of the radio source associated with the
galaxy. We can summarize our conclusions as follows: 
             
\begin{enumerate}

\item If a radio source has jets the dust tends to be in the form of 
disks or lanes, i.e. its structure is simple.

\item The radio jets tend to be perpendicular to the dust lane or disk
in FRI sources. However the effect is strong {\it only for the weaker FRIs,}
with $P < 10^{24}$~WHz$^{-1}$. In strong sources the alignment is weak or even absent.

\item Small dust masses (concentrated in small lanes or disks) are found in weak radio
sources with $P < 10^{24}$~WHz$^{-1}$.

\item This leads to one general conclusion: the weaker
FRI radio sources often have jets, which are perpendicular
to dust features that usually have smaller masses ($M/M_{\sun} \sim 10^{4.4}$) and smaller
sizes ($<500$~pc) than in stronger FRI and FRII sources; generally the dust morphology is simple (lanes and
disks). Above $10^{24}$~WHz$^{-1}$ these trends are very weak or absent, and only the following
statement holds also for the strong FRI and FRII sources:

\item There is a linear dependence of dust mass
on total (or core) radio power. Dividing the sources in different redshift intervals the correlation
remains valid also within the various redshift bins. Therefore we conclude that radio activity and the presence
of dust are strongly linked.
\end{enumerate}

\end{document}